\documentclass[aps,preprint,amsmath,amssymb]{revtex4-1}
\usepackage{graphicx}
\usepackage{here}
\usepackage{txfonts}
\usepackage{color}

\begin{document}
\title{Nonvolatile Magneto-Thermal Switching in MgB$_2$}
\author{Hiroto Arima$^1$\footnote{h-arima@tmu.ac.jp}}
\author{Yoshikazu Mizuguchi$^{1}$}
\affiliation{$^1$Department of Physics, Tokyo Metropolitan University, Hachioji, Tokyo, 192-0397, Japan}
\begin{abstract}
Ongoing research explores thermal switching materials to control heat flow. Specifically, there has been interest in magneto-thermal switching (MTS) materials based on superconductors, which only exhibited switching behavior when a magnetic field was applied. However, a recent report highlighted nonvolatile MTS in commercial Sn-Pb solders, attributed to magnetic flux trapping. In this study, we focused on flux trapping in a type-II superconductor MgB$_2$. Magnetization and thermal conductivity measurements under magnetic fields were conducted on polycrystalline MgB$_2$. We confirmed that magnetic flux was indeed trapped in MgB$_2$ even after demagnetization. Additionally, we observed nonvolatile MTS in MgB$_2$ as well as Sn-Pb solders. These results suggest that the nonvolatile MTS may be a widespread characteristic of superconducting materials with flux trapping.
\end{abstract}
\maketitle
The recent advancements in electronic device technology have spurred research into thermal switching materials, which enable control of heat flow through external parameters\cite{Nianbei2012, Wehmeyer2017}. Recent progress has been made in the development of thermal switching materials, where the control of thermal conductivity ($\kappa$) is achieved through the application of electric\cite{Ihlefeld2015} and magnetic fields\cite{Kimling2013, Hirata2023}. Among these materials, superconductors have received particular attention in magneto-thermal switching (MTS) research \cite{Yoshida_2023_Nb,yoshida_2023_Pb}. Here, we introduce an index to assess the effectiveness of MTS known as the MTS ratio (MTSR). The MTSR is calculated as the ratio of the change in $\kappa$ between the presence and absence of a magnetic field. The MTSR is expressed as [$\kappa(H)$ - $\kappa$(0 Oe)] / $\kappa$(0 Oe). It is widely recognized that, in the normal state, heat is carried by charge carriers, whereas in the superconducting state, heat transport by Cooper pairs is negligible. Consequently, the phase transition from the superconducting state to the normal state results in an increase in $\kappa$. Recent studies reported MTSR of 650 \% for Nb\cite{Yoshida_2023_Nb} and over 1000 \% for high purity 5N-Pb\cite{yoshida_2023_Pb}. However, previously reported MTS using superconductors had a limitation, $\kappa(H)$ returned to its initial value $\kappa$(0 Oe) when the magnetic field was reduced to zero, indicating that MTS was effective only in the presence of a magnetic field. In the most recent discovery reported in arXiv: 2307.05957 (preprint)\cite{Arima_2023_dolder}, a nonvolatile MTS, which retains the altered $\kappa$($H$) even when the magnetic field is completely removed, has been identified. Surprisingly, this nonvolatile MTS material was discovered in commercially available Sn-Pb solders. The nonvolatile MTSR is defined as [$\kappa$ (0 Oe, demagnetized) - $\kappa$(0 Oe, initial)]/$\kappa$ (0 Oe, initial), and it has been determined that the nonvolatile MTSR of flux-core-free Sn45-Pb55 solder was 150 \%. The origin of nonvolatile MTS in Sn-Pb solders is attributed to the presence of magnetic flux trapped in the solder even after the applied magnetic field is removed, resulting in a partial loss of superconducting bulkiness at $H = 0$ Oe. While magnetic flux trapping in Sn-Pb solders is relatively rare due to both Sn and Pb being type-I superconductors, the magnetic flux trap after demagnetization is commonly observed in type-II superconductor samples.

In this study, our primary focus is on exploring the occurrence of nonvolatile MTS in type-II superconductors, with particular emphasis on MgB$_2$, which has been studied for its flux trapping properties\cite{Mikheenko_2007,YAMAMOTO2006806}. MgB$_2$ was discovered in 2001 and stands out among intermetallic superconductors for having the highest superconducting transition temperature $T_{\rm SC}\sim39$ K under ambient pressure \cite{Nagamatsu2001}. This compound exhibits a unique characteristic as a multi-gap superconductor, with multiple conduction bands and independent superconducting gaps present on the Fermi surface\cite{Uchiyama2002,Tsuda2003}. Shortly after its discovery, it was observed that grain boundaries in MgB$_2$ could serve as effective pinning centers, contributing to high critical current density ($J_{\rm c}$) in superconducting materials\cite{Larbalestier2001, Yamamoto_2005, KATSURA2007572, takano2001}. Consequently, extensive research has been conducted to investigate the relationship between magnetic flux trapping at grain boundaries and $J_{\rm c}$.

Until now, the association between magnetic flux trapping and nonvolatile MTS has solely been reported in Sn-Pb solders. To gain a deeper understanding of this phenomenon, it is essential to explore other materials. MgB$_2$ presents an appealing platform for investigating nonvolatile MTS due to the existing body of research on flux trapping effects at grain boundaries\cite{Mikheenko_2007}. While previous studies have conducted thermal conductivity measurements under magnetic field on MgB$_2$\cite{Wu_2004,Mucha_2003}, there has been no specific focus on nonvolatile MTS. In this study, magnetization measurements and thermal conductivity measurements under magnetic fields were conducted for commercial MgB$_2$. Notably, nonvolatile MTS was also observed in MgB$_2$.

Polycrystalline MgB$_2$ used in this experiment was a commercially available powder sample (99\%, KOJUNDO). Before the measurements, the powder sample underwent a high-pressure sintering process. In this experiment, high-pressure sintering was performed at relatively low temperatures to suppress grain growth. The specific conditions for this high-pressure sintering entailed a pressure of 3 GPa and a temperature of 400 $^\circ$C, sustained around 30 minutes. The crystal structure was examined through powder X-ray diffraction employing the Cu-K$\alpha$ radiation using the $\theta$-2$\theta$ method (Miniflex-600 RIGAKU). The Rietveld refinement of the XRD data was performed using the RIETAN-FP package\cite{izumi2007}. The scanning electron microscope (SEM, TM3030, Hitachi High-Tech) was used for microstructure observation. The thermal conductivity was measured using a Physical Property Measurement System (PPMS, Quantum Design) equipped with a thermal transport option (TTO). The measurement employed a four-probe steady-state method, incorporating a heater, two thermometers, and a base-temperature terminal. For the thermal conductivity measurements of MgB$_2$, a cylindrical sample with a diameter of 4.61 mm and a height of 4.10 mm was employed. The magnetization measurements were carried out using a superconducting quantum interference device (SQUID) magnetometry technique, employing the Magnetic Property Measurement System (MPMS3, Quantum Design) in a VSM (vibrating sample magnetometry) mode. In this experiment, thermal conductivity measurements were conducted on a high-pressure sintered MgB$_2$ sample within a week. Subsequently, the sample was crushed, and further analyses including XRD and magnetization measurements, and SEM imaging were performed. All the experiments were carried out using the same batch of sample.

Figure 1 illustrates the XRD patterns obtained from the high-pressure sintered MgB$_2$ sample. In the high-pressure sintered sample, the presence of MgB$_4$ and MgO were detected as an impurity, alongside the main MgB$_2$ peaks. The reliability factor, denoted as $R_{\rm wp}$, was determined to be $R_{\rm wp}$ = 3.7 \%, and the goodness-of-fit indicator, represented by $S$, was calculated as $S$ = 1.8. The results of Rietveld refinement indicated that the sample composition consisted of approximately 90 \% MgB$_2$, 5 \% MgB$_4$, and 5\% MgO. The as-purchased MgB$_2$ powder contained a similar amount of MgB$_4$ and MgO. The discrepancy with the nominal purity of 99\% MgB$_2$ is likely a result of certain compounds not being accounted for in the chemical analysis. Furthermore, the XRD profile exhibited broadening, implying lattice strain induced by the high-pressure sintering process.

Figure 2 shows the SEM image of the high-pressure sintered MgB$_2$. Numerous granular grains were observed in the structure of the high-pressure sintered MgB$_2$, with the majority of the grain sizes measuring less than approximately 5 $\mu$m.

Figure 3 (a) illustrates the temperature dependence of the magnetization 4$\pi M$ measured at 10 Oe under both zero-field-cooling (ZFC) and field-cooling (FC) conditions. The magnetization measurement under ZFC demonstrates a large shielding signal below $T_{\rm SC}\sim$ 39 K. The difference between ZFC and FC measurements is a characteristic behavior commonly observed in type-II superconductors. The temperature dependence of 4$\pi M$ exhibited broadening, which has also been reported in previous studies on high-pressure sintered MgB$_2$\cite{takano2001}. The exact cause of this broadening is not yet clear, but the inhomogeneity of the crystals likely plays a role, as suggested by the broad profile observed in the XRD measurement. Figure 3 (b) depicts the temperature dependence of 4$\pi M$ measured at 10 Oe after FC at three different fields : 1000 Oe, 10000 Oe, and 70000 Oe. In all cases, 4$\pi M$ exhibited ferromagnetic-like behavior below $T_{\rm SC}$, similar to the findings of previously reported hydrogen-rich superconductors\cite{Minkov2023} and Sn-Pb solders\cite{Arima_2023_dolder}, implying the presence of trapped magnetic flux at grain boundaries of MgB$_2$. The value of magnetization at 1.8 K increased as the field increased from 1000 Oe to 10000 Oe, but it did not change further with the application of a higher magnetic field. This suggests that the amount of trapped magnetic flux increases with the applied magnetic field, but there is a threshold where the trapped magnetic flux saturates. To further discuss, we show the 4$\pi M$-$H$ curves at 2.5 K and 4.0 K in Figs. 3(c) and 3(e), respectively. These curves display the distinct shape commonly observed in type-II superconductors, which signifies the presence of flux trapping in the material. As depicted in Figures 3(d) and 3(f), the inner magnetic flux density ($B$) given by $B = H + 4\pi M$ near 0 Oe is displayed at 2.5 K and 4.0 K. The results at 2.5 K and 4.0 K showed similarities: immediately after the zero-field-cooling, the initial magnetic flux density of MgB$_2$ was $B$ = 0. However, upon applying a magnetic field to MgB$_2$, $B$ did not return to its initial value when the applied field reached $H$ = 0, due to the magnet flux trapping. The magnetic flux density trapped at $H$ = 0 Oe was 500 G for both temperatures.

Figure 4 (a) depicts the temperature dependence of $\kappa$ in both a zero magnetic field and a magnetic field of 10000 Oe. In the absence of a magnetic field, $\kappa$ decreased as the temperature decreased. The observed variation in the slope of $\kappa$ at approximately 10 K was consistent with previous measurements on polycrystalline MgB$_2$\cite{SCHNEIDER20016}. Furthermore, $\kappa$  at 50 K in this experiment was approximately 3.5 W/Km, which aligns with the order of magnitude reported in previous studies, where values ranged from 5 W/Km\cite{PODDAR2003191} to 9 W/Km\cite{SCHNEIDER20016}. It is noted that thermal conductivity is a sensitive indicator of grain boundaries, and therefore, the discrepancy with previous studies is attributed to the sample dependence. When a magnetic field of 10000 Oe was applied, a similar trend in $\kappa$ was observed, but the decrease in $\kappa$ was suppressed. This can be attributed to the suppression of the superconducting state in MgB$_2$ under the magnetic field. Figures 4(b) and 4(c) illustrate the magnetic field dependence of $\kappa$ at 2.5 K and 4 K, respectively. When the MgB$_2$ was zero-field-cooled to 2.5 K, the initial $\kappa$ in the absence of magnetic field was 6.9 mW/Km. When a magnetic field was applied, $\kappa$ increased and reached a value of 14.0 mW/Km at 10000 Oe. As the magnetic field gradually decreased from 10000 Oe, $\kappa$ showed a decrease. However, the value at 0 Oe deviated from the initial value, indicating nonvolatile MTS. Upon further reduction of the magnetic field, a minimum value of $\kappa$ was observed, followed by an increase in $\kappa$. Similar trends were observed when the magnetic field was increased from -10000 Oe. As mentioned earlier, the presence of approximately 500 G of trapped magnetic flux in MgB$_2$ after demagnetization partially suppresses the superconducting state and prevented $\kappa$ from returning to its initial value. The nonvolatile MTSR observed in MgB$_2$ at 2.5 K in this experiment was 18 \%, which is smaller than to that of flux-core-free Sn45-Pb55 solder\cite{Arima_2023_dolder}. Furthermore, nonvolatile MTS was also observed at 4.0 K, although the nonvolatile MTSR decreased to that at 2.5 K, reaching 15 \%.

The primary discovery of this study is the confirmation of nonvolatile MTS occurring in the magnetic flux trapped at the grain boundaries of the type-II superconductor MgB$_2$. This finding diverges from prior research, which predominantly focused on composites such as Sn-Pb solders. Notably, the phenomenon of flux trapping at grain boundaries has been observed not only in MgB$_2$ but also in other type-II superconductors, including cuprate superconductors and iron-based superconductors \cite{Hilgenkamp2002}. This suggests that the trapping of flux at grain boundaries is a widespread occurrence in various types of type-II superconducting materials. In this study, the maximum value of the nonvolatile MTSR achieved for MgB$_2$ remained relatively small at 18 \% at 2.5 K. To further enhance the nonvolatile MTSR, potential methods include controlling the grain boundary size to increase the trapped magnetic flux and regulating the thermal conductivity in the normal conducting region. However, further systematic investigations are required in this regard. Recent advancements in machine learning have contributed to the elucidation of heat conduction mechanisms in grain boundaries and nanopolycrystals \cite{Fujii2020}. Given that nonvolatile MTS is a relatively new phenomenon, it is crucial to not only investigate the thermal conductivity under magnetic field in various materials but also consider theoretical approaches that utilize machine learning to gain a deeper understanding of nonvolatile MTS.

The motivation for this study was derived from the discovery of nonvolatile MTS induced by magnetic flux trapping in Sn-Pb solders. Drawing inspiration from this phenomenon, our research focused on investigating the magnetic field dependence of thermal conductivity in type-II superconductor MgB$_2$, a material renowned for its ability to trap magnetic flux at grain boundaries. Through our experiments, we successfully observed nonvolatile MTS in MgB$_2$ and identified magnetic flux trapping as the underlying mechanism. Moving forward, it is imperative to extend this research to encompass other type-II superconductors with effective pinning centers. Such endeavors will contribute to a deeper understanding of nonvolatile MTS at a fundamental level and facilitate improvements in both the nonvolatile MTSR and the operational temperature range, thereby paving the way for potential engineering applications.

\section*{acknowledgment}
We thank O. Miura and K. Uchida for supports in experiments and fruitful discussion on the results. This work was partly supported by JST-ERATO (JPMJER2201), TMU Research Project for Emergent Future Society, and Tokyo Government-Advanced Research (H31-1).

\bibliographystyle{jpsj}
\bibliography{Reference}

\begin{figure}[h]
\begin{center}
\includegraphics[scale=0.85]{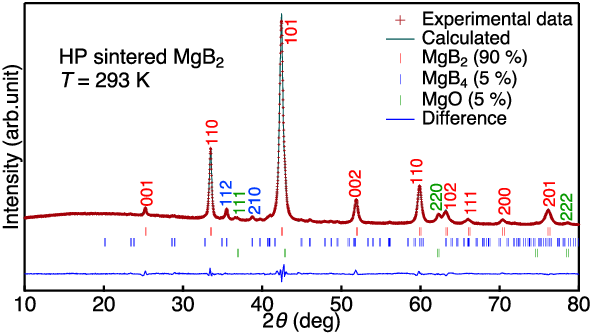}
\caption{The observed patterns ($+$ symbols), calculated patterns (green solid line), and the difference between them (blue solid line) for the high-pressure sintered MgB$_2$ at a temperature of $T$=293 K. These patterns were obtained through Rietveld refinement analysis of the XRD data. The red, blue, and green tricks represent the positions of possible Bragg reflections of MgB$_2$, MgB$_4$, and MgO, respectively. The red, blue and, green numbers indicate the Miller indices of MgB$_2$, MgB$_4$, and MgO, respectively.}
\label{struct}
\end{center}
\end{figure}

\begin{figure}[h]
\begin{center}
\includegraphics[scale=0.18]{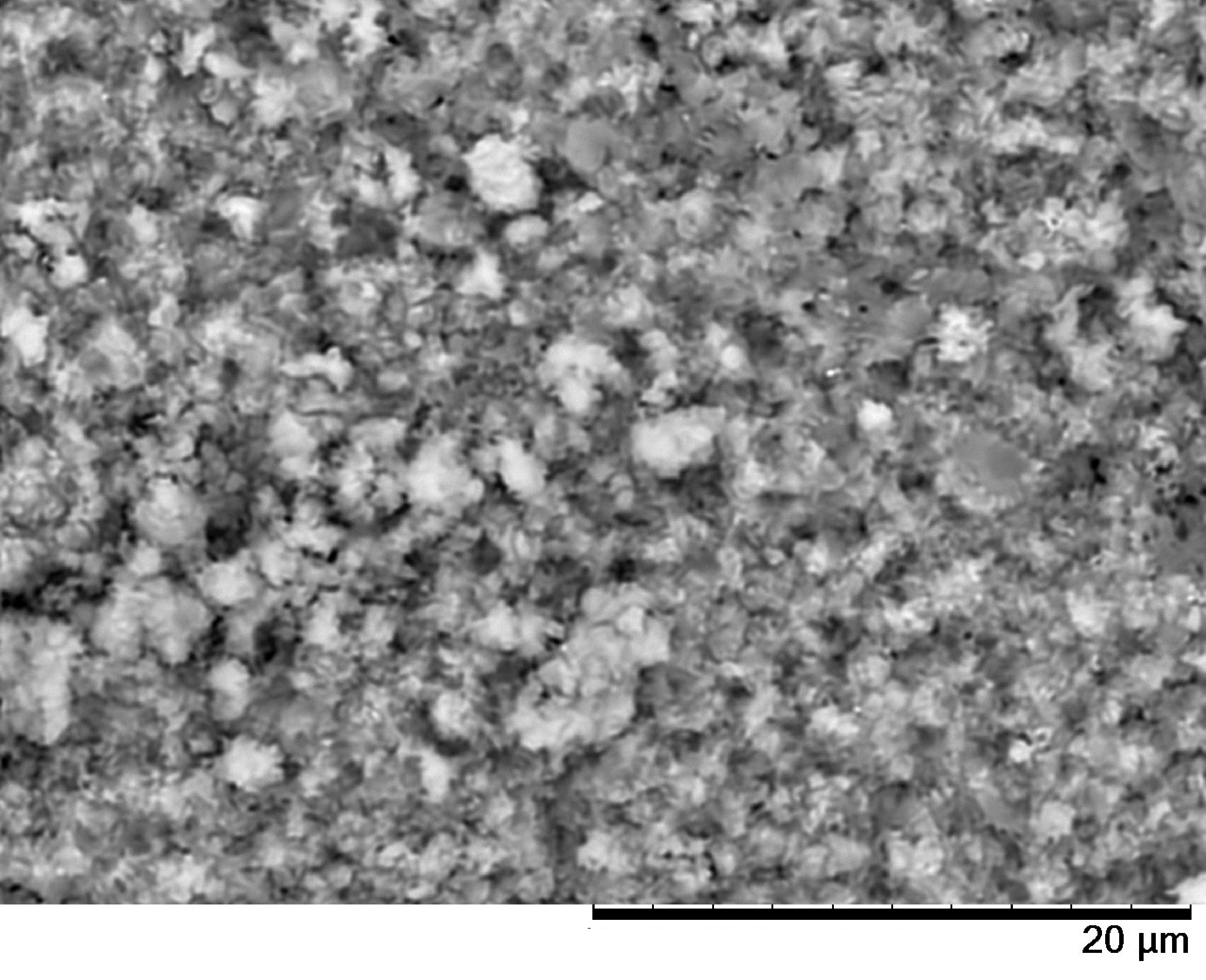}
\caption{SEM micrographs of high-pressure sintered MgB$_2$}
\label{struct}
\end{center}
\end{figure}

\begin{figure}[h]
\begin{center}
\includegraphics[scale=0.66]{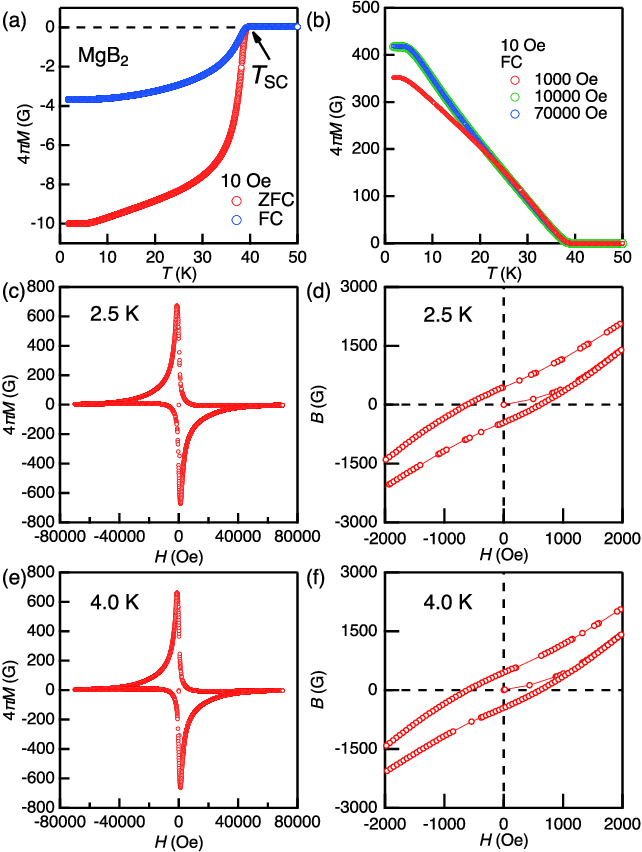}
\caption{(a) Temperature dependence of magnetization 4$\pi M$ for MgB$_2$ under conditions of zero field cooling (ZFC) and feild cooling (FC) at 10 Oe. (b) Temperature dependence of 4$\pi M$ for MgB$_2$ measured at 10 Oe after field cooling under 1000 Oe (red), 10000 Oe (green) and 70000 Oe (blue). (c-f) 4$\pi M$-$H$ and $B$-$H$ curves measured at 2.5 K and 4.0 K }
\label{struct}
\end{center}
\end{figure}

\begin{figure}[h]
\begin{center}
\includegraphics[scale=0.7]{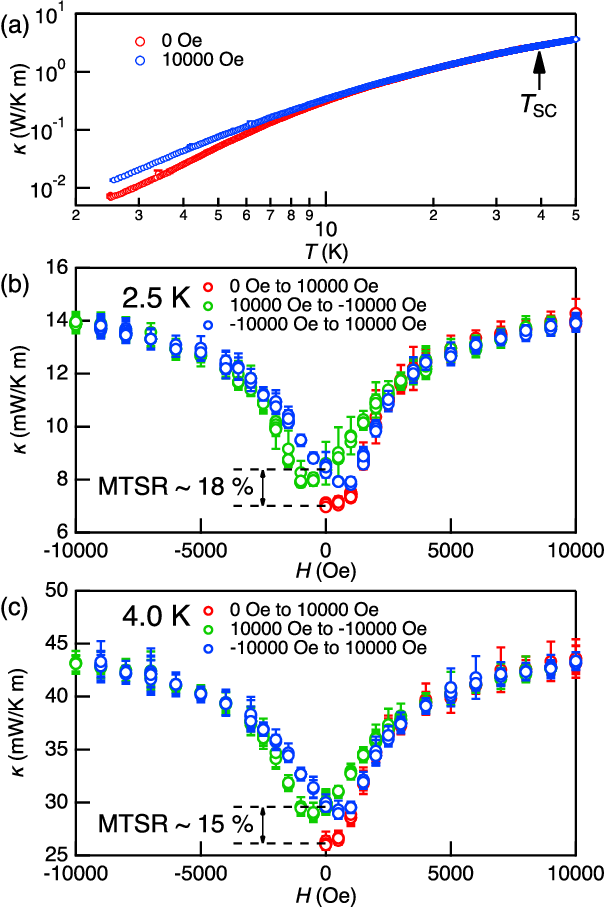}
\caption{(a) Temperature dependence of $\kappa$ measured at $H$ = 0 Oe and 10000 Oe. (b,c) $\kappa$-$H$ curves measured at $T$ = 2.5 K and 4.0 K.}
\label{struct}
\end{center}
\end{figure}

\end{document}